\documentstyle[epsfig,amssymb]{article}
\begin{document}
\title{\large\bf
Long-range correlation energies calculations for $\pi$ electronic
systems }
\author{\begin{tabular}{c} Hua Zhao \\
\begin{small}
Institute of Condensed Matter Physics and Department of Physics,
\end{small}
\\
\begin{small}
ChongQing University, ChongQing, 400044, P.R.China
\end{small}
\end{tabular}}
\date{2006/5/26}
\maketitle \footnote{E-mail:huazhao@cqu.edu.cn}
\begin{abstract}
A simple formula for correlation energy $E_c$ of the $\pi$ electron
systems is obtained under an approximation for the electron-electron
interactions. This formula is related directly to square of the bond
order matrix and the nearest-neighbor Coulomb electron-electron
interaction. The influence of the correlation energy on the band
energy gap is discussed. The values of the correlation energy for
polyacetylene (PA) are calculated and can be compared with those for
PA obtained by other methods, including $ab$ $initio$ method.
\end{abstract}
\vspace*{0.5cm}

\vspace*{0.5cm}


\section{Introduction}

The electron correlations have been a very important issue in
investigating the electronic structures of various electron systems.
Especially the electron correlations have a strong influence on the
bonding properties of atoms and semiconductor band gap$^{1}$.
However, it is well known that it is very hard to completely solve a
many electron system analytically using a single approximation
because of the exchange and correlation problems. Hartree-Fock
approximation deals with the exchange problem between the same spins
among electrons but not resolve the correlation between the opposite
spins among electrons$^{2,3}$. The local density approximation (LDA)
of Kohn and Sham$^{4}$ and later the density functional theory of
Hohenberg and Kohn$^{5}$ made a contribution to the
exchange-correlation energy, denoted by $E_{ex}[\rho(\vec{r})]$
using a complicated functional form. In the LDA, the ground state
exchange energy $E_x$ can be expressed an integral of the charge
density. However, it is difficult to express the correlation energy
$E_c$ in an available form of integral of the charge density which
is easy to calculate and only could be expressed as a numerical
formula with the parameter $r_s$ after a large number of
works$^{6}$.

Besides the LDA, there are other methods which can be used to deal
with the correlation effects of various electron systems such as
metal and nonmetal atoms, small and big molecules, as well as
polymers. They include unrestricted self-consistency field (SCF)
Approximation$^{3}$, Configuration Interaction method(CI)$^{7}$,
coupled-cluster method$^{8,9}$ that are applied to quantum chemistry
and nuclear physics$^{10}$, Jastrow wave-function method$^{11}$ that
is used to describe correlations in homogeneous fermion systems like
the electron gas or liquid He-III$^{12,13}$, the Projection
Technique$^{14}$ for the strongly correlated systems, and the GW-
approximation (the one-particle Green's function plus the screened
Coulomb interactions) by Hedin$^{15}$. For the conjugated polymers
such as polyacetylene (CH)$_x$ (PA) or poly($p$-phenylenevinylene)
(PPV), the exchange-correlation energies has been calculated by the
method of the $ab$ $initio$ with GW-approximation$^{16}$. For the
various metal and nonmetal atoms, one used atomic Bethe-Goldstone
equation under Hartree-Fock functions to calculate the electron
correlation energies for their ground states $^{17,18,19}$. For
small or medium-size molecules(hydrocarbon molecules), a
semi-empirical SCF scheme plus CNDO or INDO approximations was used
to calculate inter-atomic correlations and intra-atomic correlations
whose results could be compared with $ab$ $initio$ method$^{20}$.

However, usually the discussion about the correlation energy
concentrates mostly on the on-site Coulomb interaction, that is,
from the term $n_{i\uparrow}n_{i\downarrow}$ and seldom specially on
the lang-range Coulomb interaction, at least, on the
nearest-neighbor Coulomb interaction. For strongly correlated
systems, the on-site Hubbard interaction $U$ is much bigger than the
nearest-neighbor Coulomb interaction $v$, so the contribution of the
correlation energy are mainly from $U$. But it is well known that in
the most conjugated polymers, the on-site electron-electron
interaction $U$ is not so bigger than the nearest-neighbor
interaction $v$ because of screening$^{21}$.

In the past, people treated with the correlation energies both from
the on-site Hubbard interaction $U$ and the long-range Coulomb
interaction, In Ref.22, authors studied the correlation energies of
polyethylene (CH$_2$)$_x$(PE) using the local ansatz$^{22}$. In
Ref.23, authors used the Gutzwiller ansatz as the variational ground
state and studied correlation energy of polyacetylene (PA). First
they used the Hubbard term plus SSH Hamiltonian to discuss the
correlation energy and later added the nearest-neighbor interaction
to get an effective $U_{eff}$ and discuss the correlation energy
again. In Ref.24, author used a variational method$^{25,26}$ to
study the correlation of PA in the PPP model where both Hubbard term
and the long-range Coulomb interaction were included. But author did
not specifically deal with the correlation energy from the
long-range Coulomb interaction. It is important for the long-range
interaction to calculate the band energy gap in the conjugated
polymers$^{27}$.

Although there were those studies above about the correlation energy
both from the on-site Hubbard interaction $U$ and the long-range
Coulomb interaction and people have been known the importance of the
correlation between two $\pi$ electrons in understanding electronic
properties of the conjugated polymers, such as the optical band gap,
exciton excitation energy and exciton binding energy, polaron, etc.,
there is still lack of special studies on the correlation energy
mainly from the long-range Coulomb interaction. Therefore, study of
the correlation effect due to the Coulomb interaction of two
adjacent $\pi$ electrons in the conjugated polymers becomes
significant.

The purpose of the present work is to study the correlation energy
mainly from the long-range Coulomb interaction (the nearest-neighbor
interaction in this study). The starting point in this paper is the
exchange-correlation energy $E_{ex}$. Although the correlation
energy may be expressed as $E_c=\int\epsilon_c n(\vec{r})d\vec{r}$,
$\epsilon_c$ is hard to know and hard to obtain. Thus, available
approximations will be adopted in this study. Finally, a formula of
the electron correlation energy due to the nearest-neighbor Coulomb
interaction is obtained in an analytical form. Using this formula,
it is simple to calculated the correlation energy of a $\pi$
conjugated polymer and available to discuss the influence of the
correlation energy on the band energy gap.

The arrangement of this paper is as follows. In section II, the
expression of the pair-distribution function for the conjugated
polymer systems by omitting wave function overlap integrals between
two adjacent atomic sites is presented. In section III, an
approximation for the electron-electron interaction integral is made
and an analytical expression of the correlation energy for the
conjugated polymers is obtained. In section IV, the correlation
energy calculation for one-dimensional polyacetylene (PA) chain
under tight-bind approximation (SSH Hamiltonian plus
electron-electron interactions), and results are presented, and in
section V is discussion with a summary.

\section{Pair-distribution function for electron systems}

The so-called correlation is the correlation hole appearing around
an electron moving in the medium. Remarkably the correlation in fact
is due to the electron-electron interaction among electrons. In a
homogeneous or an inhomogeneous electron gas, the density operator
of the electrons is expressed as a delta function
$\hat{\rho}(\vec{r})=\sum_{i}\delta(\vec{r}-\vec{r}_i)$, the density
distribution function of the electrons is the average of the
operator over a given state $|\Phi>$, $\rho(\vec{r})
=<\Phi|\hat{\rho}(\vec{r})|\Phi>$. The two-point density-density
correlation function at a given state $|\Phi>$ is defined as
\begin{eqnarray}
C(\vec{r},\vec{r'})&=&
 <\Phi|\sum_{i\neq j}\delta(\vec{r'}-\vec{r}_i)
 \delta(\vec{r}-\vec{r}_j)|\Phi>
 \nonumber \\
 &\equiv &<\hat{\rho}(\vec{r})\hat{\rho}(\vec{r'})>
 \label{eq:Corr}
\end{eqnarray}
where $<\dots>$ means $<\Phi|\dots|\Phi>$, the average of the
operators over the ground states. To calculate the two-point
density-density correlation function, one introduces the
pair-distribution function which is defined as$^{28}$
\begin{eqnarray}
 <\hat{\rho}(\vec{r})\hat{\rho}(\vec{r'})>
 =g(\vec{r},\vec{r'})<\hat{\rho}(\vec{r})><\hat{\rho}(\vec{r'})>
 \label{eq:g}
\end{eqnarray}
About the pair-distribution function $g(\vec{r},\vec{r'})$ there are
different approximations. For example it can be cast into an
analytic expression for the homogeneous gas$^{29}$. For a
many-electron system, the density distribution $\rho(\vec{r})$ of
the electrons can be expressed by
\begin{eqnarray}
\rho(\vec{r})&=&2\sum_{\mu}^{occ}|\psi_{\mu}(\vec{r})|^2
\nonumber \\
&=& \sum_{ij}^N\rho_{ij}\phi^*_i(\vec{r})\phi_j(\vec{r})
\label{eq:Density}
\end{eqnarray}
where the index $\mu$ refers to the occupied molecular orbital,
$\mu=1,2,\cdots, N/2$. The indices $i$ and $j$ refer to the atomic
sites. $\phi_i(\vec{r})$ is the electron orbital wave function at
the atomic site $i$ and is normalized to one. Here
$\rho_{ij}=\sum_{\sigma}<a^{\dag}_{i\sigma}a_{j\sigma}>$ where
$a^{\dag}_{i\sigma}$ ($a_{j\sigma}$) is the electron creative
(annihilation) operator with spin $\sigma$ at the site $i$ ($j$).
$\rho_{ij}$ is called the bond-order matrix$^{28}$. Comparing
(\ref{eq:Density}) with $\rho(\vec{r})
=<\Phi|\hat{\rho}(\vec{r})|\Phi>$, we can see that the density
operator for a many-electron system can be represented by
\begin{eqnarray}
\hat{\rho}(\vec{r})=
\sum_{i,j}^N\hat{\rho}_{ij}\phi^*_i(\vec{r})\phi_j(\vec{r})
\label{eq:rho-D}
\end{eqnarray}
with $\hat{\rho}_{ij}=\sum_{\sigma}a^{\dag}_{i\sigma}a_{j\sigma}$.
Then the two-point density-density correlation function is expressed
as
\begin{eqnarray}
<\hat{\rho}(\vec{r})\hat{\rho}(\vec{r'})> &=& \sum_{\sigma\sigma'}
\sum_{ij}^N\sum^N_{kl}<\hat{\rho}^{\sigma}_{ij}
 \phi^*_i(\vec{r})\phi_j(\vec{r})
 \hat{\rho}^{\sigma'}_{kl}\phi^*_k(\vec{r'})\phi_l(\vec{r'}) >
 \nonumber \\
 &=&
\sum_{\sigma\sigma'}\sum_{ik(i\neq k)}^N<\hat{\rho}^{\sigma}_{ii}
 \hat{\rho}^{\sigma'}_{kk}>
 |\phi_i(\vec{r})|^2|\phi_k(\vec{r'})|^2
 \nonumber \\
 &+&
 \sum_{\sigma\sigma'}\sum_{ij(i\neq j)}^N
 \sum^N_{kl(k\neq l)}<\hat{\rho}^{\sigma}_{ij}
 \hat{\rho}^{\sigma'}_{kl}> \phi^*_i(\vec{r})\phi_j(\vec{r})
 \phi^*_k(\vec{r'})\phi_l(\vec{r'})
 \label{eq:den-den}
\end{eqnarray}
and
\begin{eqnarray}
<\hat{\rho}(\vec{r})> & = &
<\sum_{ij\sigma}^N\hat{\rho}^{\sigma}_{ij}\phi^*_i(\vec{r})
\phi_j(\vec{r})> \nonumber \\
&=& \sum_{\sigma i}<\hat{\rho}^{\sigma}_{ii}> |\phi_i(\vec{r})|^2
+\sum_{\sigma}\sum_{i\neq j}^N <\hat{\rho}^{\sigma}_{ij}>
\phi^*_i(\vec{r})\phi_j(\vec{r})
 \label{eq:oper-rho}
\end{eqnarray}
A calculation containing the pair-distribution function involves
wave function integrals between different sites, which is related to
the overlap of two electronic wave functions. As a first order
approximation and without loss generality, the overlap are not
considered, so the second terms in (\ref{eq:den-den}) and
(\ref{eq:oper-rho}) could be omitted in this study. Therefore, the
pair-distribution function $g(\vec{r},\vec{r'})$ in (\ref{eq:g})
becomes
\begin{eqnarray}
g_{\sigma\sigma'}(\vec{r},\vec{r'}) &=& \frac{\sum_{ik(i\neq
k)}^N<\hat{\rho}^{\sigma}_{ii}
 \hat{\rho}^{\sigma'}_{kk}>
 |\phi_i(\vec{r})|^2|\phi_k(\vec{r'})|^2}
 {\sum_{\sigma i}<\hat{\rho}^{\sigma}_{ii}> |\phi_i(\vec{r})|^2
\sum_{\sigma k}<\hat{\rho}^{\sigma}_{kk}> |\phi_k(\vec{r'})|^2}
 \label{eq:g-3}
\end{eqnarray}
with $g(\vec{r},\vec{r'})=\sum_{\sigma\sigma'}g(\vec{r},\vec{r'})$
and $\rho^{\sigma}_{ij}=<a^{\dag}_{i\sigma}a_{j\sigma}>$.

\section{Approximation approach to the correlation energy}

In the local density approximation (LDA) the exchange-correlation
energy can be written, in terms of the pair-distribution function
$\widetilde{g}(\vec{r},\vec{r'})$ in the form$^{30,31}$
\begin{eqnarray}
E_{xc}[\rho] =\frac{e^2}{2}\int d^3r d^3r' \rho(\vec{r})
\frac{[\widetilde{g}(\vec{r},\vec{r'})-1]}{|\vec{r}-\vec{r'}|}
\rho(\vec{r'}) \label{eq:Exc}
\end{eqnarray}
where
\begin{eqnarray}
\widetilde{g}(\vec{r},\vec{r'})=\int^1_0 d\lambda
g(\vec{r},\vec{r'};\lambda). \label{eq:g-lambda}
\end{eqnarray}
Here $\widetilde{g}(\vec{r},\vec{r'})$ includes the
exchange-correlation information between two spatial points
$\vec{r}$ and $\vec{r'}$. Here $\lambda$ means the charge $e^2$ in
Coulomb interaction is replaced by $\lambda e^2$ in the process of
calculation. In (\ref{eq:Exc}),
$\rho(\vec{r'})[\tilde{g}(\vec{r},\vec{r'})-1]$ describes the
depletion in density due to the exchange-correlation hole around an
electron at position $\vec{r}$. The density depletion due to the
exchange-correlation hole around an electron corresponds to exactly
one particle, independent of the size of the electron-electron
coupling strength. Then we have the sum rule satisfied by the
exchange-correlation hole:
\begin{eqnarray}
\int d^3\vec{r'}\rho(\vec{r'})[\tilde{g}(\vec{r},\vec{r'})-1] =-1.
\label{eq:sum-rule}
\end{eqnarray}
It is well known that it is rather difficult to compute the
integration in the relation (\ref{eq:Exc}) analytically due to the
pair-distribution function $\tilde{g}(\vec{r},\vec{r'})$.

Since the HF approximation has already contained the contribution
from the exchange effect between two electrons with the same spins,
the correlation energy may be obtained by subtracting the HF
pair-distribution function $g^{HF}_{\sigma\sigma}(\vec{r},\vec{r'})$
from the exchange-correlation energy $E_{xc}$$^{28,32}$,
\begin{eqnarray}
E_c[\rho_{\uparrow}, \rho_{\downarrow} ]
=\frac{e^2}{2}\sum_{\sigma\sigma'}\int d^3r d^3r'
\frac{\rho_{\sigma}(\vec{r})
\rho^c_{\sigma\sigma'}(\vec{r},\vec{r'})}
{|\vec{r}-\vec{r'}|} \label{eq:Ec}
\end{eqnarray}
where $\rho^c_{\sigma\sigma'}$ is the spin-dependent correlation
hole and given by
\begin{eqnarray}
\rho^c_{\sigma\sigma'}(\vec{r},\vec{r'})=\rho_{\sigma'}(\vec{r'})
[\widetilde{g}_{\sigma\sigma'}(\vec{r},\vec{r'})
-\delta_{\sigma\sigma'}\widetilde{g}^{HF}_{\sigma\sigma'}
(\vec{r},\vec{r'})].
\label{eq:rho-c}
\end{eqnarray}
where $\widetilde{g}^{HF}_{\sigma\sigma'}(\vec{r},\vec{r'})$ is the
pair-distribution function under the HF approximation, and is given
by
\begin{eqnarray}
\widetilde{g}^{HF}_{\sigma\sigma'}(\vec{r},\vec{r'})=\int^1_0
d\lambda g^{HF}_{\sigma\sigma'}(\vec{r},\vec{r'};\lambda).
\label{eq:g-HF}
\end{eqnarray}
The difference between the two pair-distribution functions
$(\widetilde{g}-\widetilde{g}^{HF})$ has to do with the electron
correlation. The expression (\ref{eq:Ec}) now can be regarded as an
integral between the $\pi$ electron at the position $\vec{r}$ and
the charge cloud of the spin-dependent correlation hole around the
position $\vec{r'}$ of the other $\pi$ electron. Thus the expression
(\ref{eq:Ec}) can be rewritten as
\begin{eqnarray}
E_c =\frac{1}{2}\sum_{\sigma\sigma'}\int d^3r d^3r'
\rho_{\sigma}(\vec{r})v(\vec{r}-\vec{r'})
\rho^c_{\sigma\sigma'}(\vec{r},\vec{r'})
 \label{eq:Ec-rr}
\end{eqnarray}
where $v(\vec{r}-\vec{r'})=\frac{e^2}{|\vec{r}-\vec{r'}|}$, the
Coulomb interaction between two electrons at the positions $\vec{r}$
and $\vec{r'}$.

In order to complete the calculation of the integration of $E_c$, an
approximation is needed to simplify the expression for $E_c$. In the
study of polyacetylene (PA) oligomer, two-electron interaction
integral was approximated by abstracting the Coulomb interaction
$v(\vec{r}-\vec{r'})$ out of the integrand of the many-centered
Coulomb integral, and the results showed that the approximation is
reasonable$^{33}$. According to that spirit, we may take
$v(\vec{r}-\vec{r'})$ out of the integrand in (\ref{eq:Ec-rr}). For
simplicity, in this study only consider the nearest-neighbor
electron-electron interaction between two adjacent carbon atom sites
are considered, that is, $v(\vec{r}-\vec{r'})=\lambda v$, where
$\lambda$ is due to the replacement of the charge $e^2$ in the
Coulomb interaction. In this way the expression for $E_c$ may become
\begin{eqnarray}
 E_c &\approx &\frac{\lambda v}{2}\sum_{\sigma\sigma'}\int d^3r d^3r'
 \rho_{\sigma}(\vec{r})\rho^c_{\sigma\sigma'}(\vec{r},\vec{r'})
 \nonumber\\
 &=& \frac{\lambda v}{2}\sum_{\sigma\sigma'}\int d^3r d^3r'
 \rho_{\sigma}(\vec{r}) \rho_{\sigma'}(\vec{r'})
 [\widetilde{g}_{\sigma\sigma'}(\vec{r},\vec{r'})
 -\delta_{\sigma\sigma'}\widetilde{g}^{HF}_{\sigma\sigma'}
 (\vec{r},\vec{r'})]
 \nonumber \\
 &=& \frac{\lambda v}{2}[\sum_{\sigma\sigma'}\int d^3r d^3r'
 \rho_{\sigma}(\vec{r}) \rho_{\sigma'}(\vec{r'})
 \widetilde{g}_{\sigma\sigma'}(\vec{r},\vec{r'})
 -\sum_{\sigma}\int d^3r d^3r'
 \rho_{\sigma}(\vec{r}) \rho_{\sigma}(\vec{r'})
 \widetilde{g}^{HF}_{\sigma\sigma}(\vec{r},\vec{r'})]
 \label{eq:Ec-rr-2}
\end{eqnarray}
Remarkably, it is not easy to calculate it without further
approximation. According to the sum rule (\ref{eq:sum-rule}), we
have
\begin{eqnarray}
\sum_{\sigma\sigma'}\int d^3r d^3r'
 \rho_{\sigma}(\vec{r}) \rho_{\sigma'}(\vec{r'})
 \widetilde{g}_{\sigma\sigma'}(\vec{r},\vec{r'})
 =\frac{1}{2}\sum_{\sigma\sigma'}\int d^3r d^3r'
 \rho_{\sigma}(\vec{r}) \rho_{\sigma'}(\vec{r'})
 -\frac{1}{2}\sum_{\sigma}\int d^3r
 \rho_{\sigma}(\vec{r})
 \label{eq:sum-rule-2}
\end{eqnarray}
Then the expression for $E_c$ can be reduced to
\begin{eqnarray}
E_c &=& \frac{\lambda v}{2}[-\frac{1}{2}\sum_{\sigma}\int d^3r
\rho_{\sigma}(\vec{r}) + \frac{1}{2}\sum_{\sigma\sigma'}\int d^3r
d^3r' \rho_{\sigma}(\vec{r})\rho_{\sigma' }(\vec{r'}) -
\sum_{\sigma}\int d^3r d^3r'
\rho_{\sigma}(\vec{r})\rho_{\sigma}(\vec{r'})
\widetilde{g}^{HF}_{\sigma\sigma}(\vec{r},\vec{r'})].
 \label{eq:Ec-4}
\end{eqnarray}

Now the remaining central task is to calculate the Hartree-Fock
pair-distribution function $g^{HF}(\vec{r},\vec{r}')$. From
(\ref{eq:g-3}), we have
\begin{eqnarray}
g^{HF}_{\sigma\sigma}(\vec{r},\vec{r'}) &=& \frac{\sum_{ik(i\neq
k)}^N<\Phi_{HF}|\hat{\rho}^{\sigma}_{ii}\hat{\rho}^{\sigma}_{kk}
|\Phi_{HF}> |\phi_i(\vec{r})|^2|\phi_k(\vec{r'})|^2}
{\sum^N_{i\sigma}<\Phi_{HF}|\hat{\rho}^{\sigma}_{ii}|\Phi_{HF}>
|\phi_i(\vec{r})|^2
\sum^N_{k\sigma}<\Phi_{HF}|\hat{\rho}^{\sigma}_{kk}|\Phi_{HF}>
|\phi_k(\vec{r'})|^2}.
 \label{eq:g-HF-2}
\end{eqnarray}
where $|\Phi_{HF}>$ is the HF ground state.

Inserting (\ref{eq:Density}) and (\ref{eq:g-HF-2}) into the
expression of $E_c$, it yields
\begin{eqnarray}
E_c &=& \frac{\lambda
v}{2}[-\frac{1}{2}\sum_{i\sigma}\rho^{\sigma}_{ii} +
\frac{1}{2}\sum^{i\neq k}_{ik,\sigma\sigma'}
\rho^{\sigma}_{ii}\rho^{\sigma'}_{kk}
 - \sum^{i\neq k}_{ik,\sigma}
<\Phi_{HF}|\hat{\rho}^{\sigma}_{ii}\hat{\rho}^{\sigma}_{kk}
|\Phi_{HF}>].
 \label{eq:Ec-5}
\end{eqnarray}
The third term in the above expression can be evaluated and it
equals
\begin{eqnarray}
\sum_{\sigma}
<\Phi_{HF}|\hat{\rho}^{\sigma}_{ii}\hat{\rho}^{\sigma}_{kk}
|\Phi_{HF}>
&=&\sum_{\sigma\sigma'}<\Phi_{HF}|a^{\dag}_{i\sigma}a_{i\sigma}
a^{\dag}_{k\sigma}a_{k\sigma}|\Phi_{HF}>
\nonumber\\
&=&\sum_{\sigma}[<a^{\dag}_{i\sigma}a_{i\sigma}>
<a^{\dag}_{k\sigma}a_{k\sigma}> -<a^{\dag}_{i\sigma}a_{k\sigma}>
<a^{\dag}_{k\sigma}a_{i\sigma}>]
\nonumber\\
&=&\sum_{\sigma}[\rho^{\sigma}_{ii} \rho^{\sigma}_{kk}
-(\rho^{\sigma}_{ik})^2].
 \label{eq:HF}
\end{eqnarray}
Here the symmetry is used that $<a^{\dag}_{i\sigma}a_{k\sigma}>
=<a^{\dag}_{k\sigma}a_{i\sigma}>$. Finally the correlation energy
obtained is
\begin{eqnarray}
E_c &=& \frac{\lambda
v}{2}[-\frac{1}{2}\sum_{i\sigma}\rho^{\sigma}_{ii}+ \sum^{i\neq
k}_{ik\sigma} (\rho^{\sigma}_{ik})^2].
 \label{eq:Ec-6}
\end{eqnarray}
This expression (\ref{eq:Ec-6}) is the central result in this paper
and it gives the correlation energy for an open or closed
one-dimensional $\pi$ electronic system in real space. The
correlation energy per electron is $\epsilon_c=E_c/N$, $N$ is the
total numbers of the carbon atoms in a $\pi$ electron system. For a
half filled system, the number of atomic sites and the number of
electrons are equal. This relation tells us that when the averages
of the bond charge and the electron density at the site $i$ are
known, the correlation energy can be evaluated.

\section{Calculation and results}

In this study, the main consideration is about the correlation
energy of the $\pi$ electron systems. For the $\pi$ conjugated
polymers, the Hamiltonian of the system is the SSH-type Hamiltonian,
$H_0$, plus the electron-electron interaction term,
\begin{eqnarray}
H=H_0+\frac{1}{2}\sum_{ij\sigma\sigma'}v(\vec{r}-\vec{r'})
a^{\dag}_{i\sigma}a_{i\sigma}a^{\dag}_{j\sigma'}a_{j\sigma'}
 \label{eq:H}
\end{eqnarray}
\begin{eqnarray}
H_0=\sum_{ij\sigma\sigma'}t_{ij}(a^{\dag}_{i\sigma}a_{j\sigma}+
h.c.)
 \label{eq:H-0}
\end{eqnarray}
where $a^{\dag}_{i\sigma}$($a_{j\sigma}$) is the creation
(annihilation) operator of an $\pi$ electron at the site $i$ ($j$)
with spin $\sigma$. $v(\vec{r}-\vec{r'})$ is the electron-electron
interaction, and $\vec{r}$ ($\vec{r'}$) means the position vector of
an $\pi$ electron at the site $i$ ($j$). $t_{ij}$ is the hopping
term. For an bond-alternated chain, $t_{i,i+1}= t_0+(-1)^i\delta
t_0$ with $t_0$ being the hopping integral without dimerization and
$\delta t_0$ being the magnitude of the dimerization due to Peierls
transition.

For the one-dimensional $\pi$ electron conjugated polymers with N
carbon atoms such as the bond-alternated chain PA, in the bond order
wave (BOW) phase, the average charge density at the site $i$
$\rho^{\sigma}_{ii}=1/2$, and the average of the bond charge density
$\rho^{\sigma}_{ik}=\rho^{\sigma}_{ki}=
<a^{\dag}_{i\sigma}a_{k\sigma}>$ (here $k=i+1$). Dropping the spin
index because of
$\rho^{\sigma}_{ik}=\rho^{\bar{\sigma}}_{ik}=\rho_{ik}$, we have
$\rho_{ii+1}=\bar{\rho}+(-1)^i\delta\rho$. Then (\ref{eq:Ec-6})
becomes
\begin{eqnarray}
E_c &=& \frac{\lambda v}{2} [-\frac{N}{2} + 2\sum_i\rho^2_{i,i+1}]
 \nonumber\\
&=& \frac{\lambda v}{2} [-\frac{N}{2} +
2\sum_{m=1}^{N/2}(\rho^2_{2m-1,2m}+\rho^2_{2m,2m+1})].
 \label{eq:Ec-7}
\end{eqnarray}
Since $\rho^2_{2m-1,2m}=(\bar{\rho}-\delta\rho)^2= \bar{\rho}^2
-2\bar{\rho}\delta\rho +(\delta\rho)^2$ and
$\rho^2_{2m,2m+1}=(\bar{\rho}+\delta\rho)^2= \bar{\rho}^2
+2\bar{\rho}\delta\rho +(\delta\rho)^2$, so the correlation energy
for PA may be expressed as
\begin{eqnarray}
E_c&=& -N\frac{\lambda v}{2}[\frac{1}{2} - 2(\bar{\rho}^2 +
(\delta\rho)^2)].
 \label{eq:Ec-8}
\end{eqnarray}
where $N$ is the carbon atom numbers in the system. The correlation
energy per $\pi$ electron is then given by the integral over the
parameter $\lambda$ from 0 to 1,
\begin{eqnarray}
\epsilon_c&=& -\frac{v}{2}\{\int_0^1\frac{\lambda}{2}d\lambda -
2\int_0^1[\lambda(\bar{\rho}(\lambda))^2 +
\lambda(\delta\rho(\lambda))^2]d\lambda \}.
 \label{eq:Ec-9}
\end{eqnarray}
Here $\bar{\rho}(\lambda)$ and $\delta\rho(\lambda)$ are given by
the first and second elliptic integrals$^{34}$
\begin{eqnarray}
\bar{\rho}(\lambda)=\frac{1}{\pi(1-z(\lambda)^2)}
[E(\sqrt{1-z(\lambda)^2})-z(\lambda)^2K(\sqrt{1-z(\lambda)^2})]
\label{eq:first-elliptic}
\end{eqnarray}
\begin{eqnarray}
\delta\rho(\lambda)=\frac{z(\lambda)}{\pi(1-z(\lambda)^2)}
[K(\sqrt{1-z(\lambda)^2})-E(\sqrt{1-z(\lambda)^2})]
\label{eq:second-elliptic}
\end{eqnarray}
where $z(\lambda)=\delta t/t$ and the parameters $t$ and $\delta t$
are determined by the relations
\begin{eqnarray}
t=t_0+ \lambda v\bar\rho_0
 \label{eq:t-pa}
\end{eqnarray}
\begin{eqnarray}
\delta t=\delta t_0+ \lambda v\delta\rho_0
 \label{eq:dt-pa}
\end{eqnarray}
where $\bar\rho_0$ and $\delta\rho_0$ are the average charge density
and the change of the charge density without the Coulomb
interactions.

For the PA, $t_0=$2.5 eV and $\delta t_0=2\alpha u_0=0.269$ eV,
where $\alpha=4.1$ eV$/{\AA}$ and the dimerization
$u_0=0.0328$${\AA}$$^{27}$. In calculation, the lang-range
interaction $v$ is 2.4 eV$^{27}$. Table 1 lists the values for
different $\lambda$ when using the the relations (\ref{eq:t-pa}) and
(\ref{eq:dt-pa}) to calculate the average of the bond charge density
$\bar{\rho}$ and its change $\delta\rho$ from the relations
(\ref{eq:first-elliptic}) and (\ref{eq:second-elliptic}). Putting
the resulting $\bar{\rho}$ and $\delta\rho$ into the (\ref{eq:Ec-9})
and integrating over $\lambda$ from 0 to 1, the correlation energy
is then obtained. In Table 1
$I(\lambda)=(\bar{\rho}(\lambda))^2+(\delta\rho(\lambda))^2$, and
$I=2\int_0^1[\lambda(\bar{\rho}(\lambda))^2 +
\lambda(\delta\rho(\lambda))^2]d\lambda $. Table 2 lists the values
of the correlation energies for PE and PA for contributions from the
on-site Coulomb interaction and the nearest-neighbor Coulomb
interaction, and also lists the band energy gap $E'_g$ in this study
and that from Ref.27. Fig.1 shows the correlation energy vs the
dimerization $\delta t_0$.

To see the influence of the parameter $\lambda$ on the the bond
charge density and finally on the correlation energy, I calculate
the $\bar{\rho}(\lambda)$ and $\delta\rho(\lambda)$ (see the lines
from 2nd to 11th in the Table 1). It is seen from the Table 1 that
the differences about the various quantities when $\lambda=0$ and
when $\lambda=1$ are very small:
$\bar{\rho}(0)-\bar{\rho}(1)=0.0026$,
$|\delta\rho(0)-\delta\rho(1)|=0.0200$, $|I(0)-I(1)|=0.0024$. In
addition, the differences between $I$ and $I(\lambda)$ are also
small: $|I-I(0)|=0.0017$ eV, $|I-I(1)|=0.0007$ eV, and
$|\epsilon_c-\epsilon_c(\lambda=0)|=0.0020$ eV,
$|\epsilon_c-\epsilon_c(\lambda=1)|=0.0009$ eV.

The treatment of the pair-distribution function (\ref{eq:g-lambda}),
$\widetilde{g}(\vec{r},\vec{r}')=\int^1_0 d\lambda
g(\vec{r},\vec{r'};\lambda)$ is as follows. In the formula
(\ref{eq:g-lambda}), Coulomb interaction $e^2$ in the Coulomb
interaction $v(\vec{r}-\vec{r'})$ is replaced by $\lambda e^2$,
$\lambda$ changes from 0 to 1. When $v(\vec{r}-\vec{r'})$ appears
and changes, the electron wave functions follows the changes. Then
the electron density distribution $\rho(\vec{r})$ changes, and then
$\rho_{ij}$ changes, which may be expressed as $\bar{\rho}(\lambda)$
and $\delta\rho(\lambda)$ (see (\ref{eq:first-elliptic}) and
(\ref{eq:second-elliptic})). This can be seen through the
expressions (\ref{eq:t-pa}) and (\ref{eq:dt-pa}). Therefore the
calculation of $\widetilde{g}(\vec{r},\vec{r}')$ is realized
actually through the calculation of $\rho(\vec{r})(\lambda)$. Table
1 shows the results of the correlation energies with integration
about the parameter $\lambda$ (see the first line in the Table 1).

Note that the bond charge density $\rho^{\sigma}_{ij}$ always is
less than half and about $0.3$ or so, therefore the second term is
less than the first term in the formula (\ref{eq:Ec-6}) and the
correlation energy is negative. The electron systems to which the
formula is suitable is supposed to be $\pi$ conjugated polymers with
long chain ($N$ is very large) with half filled band in the ground
state. For a small molecule system such as $H_2$, etc, and those
without $\pi$ electrons, the formula is not suitable because the
molecule like $H_2$ is covalent molecules where the charge density
gathers between two atoms. Another reason is that the approximation
(\ref{eq:Ec-rr-2}) may bring a bigger error about the
electron-electron interaction integral if it is applied to small
molecules like $H_2$ etc. It is also not suitable to the hydrocarbon
such as methane molecule that has no $\pi$ electrons. In derivation
of the formula, there is no the excited states to be dealt with.

\begin{table}
  \centering
  \caption{ To see the influence of the parameter $\lambda$ on the
the bond charge density and finally on the correlation energy, the
$\bar{\rho}(\lambda)$ and $\delta\rho(\lambda)$ are calculated (see
the lines from 2nd to 11th in the Table). It is seen that the
differences about the various quantities when $\lambda=0$ and when
$\lambda=1$ are very small.
  $t_0=2.5$, $u_0=0.0328$,$\alpha=4.1 eV/\AA$
  and $v=2.4$ for PA. The unit is eV.}\label{90}
 \begin{tabular}{cccccccccc}
  \hline\hline
  $z_0$ & $\bar{\rho}_0$ & $\delta\rho_0$ & $\lambda$ & z &
  $\bar{\rho}(\lambda)$ & $\delta\rho(\lambda)$ & $I(\lambda)$
  & $I$ & $\epsilon_c$   \\
  \hline \\
   0.1076 & 0.3144 & 0.0903 &&&&&& 0.1087 & -0.1696   \\
   \hline \\
   0.1076 & 0.3144 & 0.0903 &  0  & 0.1076 & 0.3144 & 0.0903 &
   0.1070 &     & -0.1716   \\
   &  &   & 0.1 & 0.1129 & 0.3141 & 0.0930 &
   0.1073 &     & -0.1713   \\
   &  &   & 0.2 & 0.1178 & 0.3138 & 0.0956 &
   0.1076 &     & -0.1709   \\
   &  &   & 0.3 & 0.1225 & 0.3135 & 0.0979 &
   0.1079 &     & -0.1706   \\
   &  &   & 0.4 & 0.1269 & 0.3132 & 0.1001 &
   0.1081 &     & -0.1703   \\
   &  &   & 0.5 & 0.1311 & 0.3130 & 0.1021 &
   0.1084 &     & -0.1700   \\
   &  &   & 0.6 & 0.1351 & 0.3127 & 0.1039 &
   0.1086 &     & -0.1697   \\
   &  &   & 0.7 & 0.1389 & 0.3125 & 0.1057 &
   0.1088 &     & -0.1694   \\
   &  &   & 0.8 & 0.1425 & 0.3122 & 0.1073 &
   0.1090 &     & -0.1692   \\
   &  &   & 0.9 & 0.1460 & 0.3120 & 0.1089 &
   0.1092 &     & -0.1690  \\
   &  &   & 1.0 & 0.1492 & 0.3118 & 0.1103 &
   0.1094 &     & -0.1687  \\
  \hline\hline
 \end{tabular}
\end{table}

\begin{table}
  \centering
  \caption{Correlation energies $\epsilon_c$ from $n_in_j$
  and $n_{i\uparrow}n_{i\downarrow}$ for PA and PE.
  Here $\epsilon_c$(V) means contribution from the
  nearest-neighbor Coulomb interaction, $\epsilon_c$(U)
  means contribution from the on-site Hubbard interaction,
  and $\epsilon_c$(U+V)means contribution from both the
  on-site Hubbard interaction and the nearest-neighbor
  Coulomb interaction. $E'_g (=E_g+\epsilon_c$) and
  $E''_g$ are the band energy gap. The unit is eV.}\label{95}
 \begin{tabular}{ccc}
  \hline\hline
  Systems  & PE & PA  \\
    \hline \\
  $E'_g$          &              &  1.7832$^*$    \\
  $E''_g$         &              &  1.8 $^d$      \\
    \hline \\
 $\epsilon_c$(V)   & -0.1725$^a$ & -0.1696$^*$  \\
 $\epsilon_c$(V)   &             & -0.1567$^b$  \\
 $\epsilon_c$(V)   &             & -0.0100$^c$  \\
 $\epsilon_c$(U)   & -0.5760$^a$ &    \\
 $\epsilon_c$(U)   &             & -0.7885$^b$  \\
 $\epsilon_c$(U)   &             & -0.6208$^c$  \\
 $\epsilon_c$(U+V) & -0.75$^a$   &     \\
 $\epsilon_c$(U+V) &             & -0.6319$^b$  \\
 $\epsilon_c$(U+V) &             & -0.6308$^c$  \\
  \hline\hline
   $a$ refers to the reference[22]\\
   $b$ refers to the reference[23]\\
   $c$ refers to the reference[24]\\
   $d$ refers to the reference[27]\\
   $*$ refers to the present study\\
 \end{tabular}
\end{table}

\index{Fig}\index{PostScript}
\begin{figure}[hbt]
 \epsfig{file=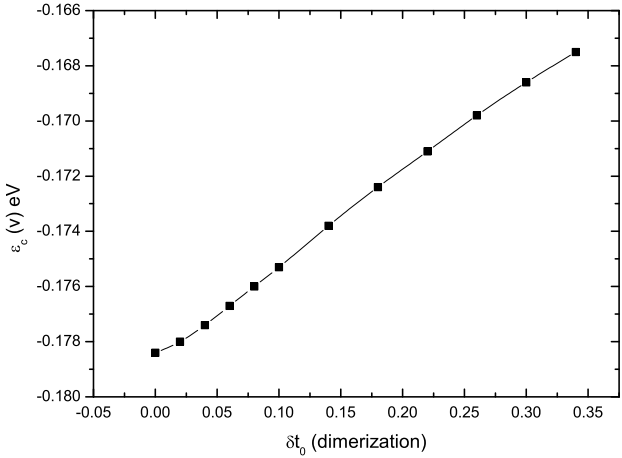, height=3in}
 \caption{ $\epsilon_c$ changes with increasing
 dimerization $\delta t_0$ ($=2\alpha u_0$) under $v=2.4$ eV
 according to the formula (\ref{eq:Ec-9}).
 Here $v$ in $\epsilon_c(v)$ means the nearest-neighbor
 Coulomb interaction. }
\label{Fpsfig}
\end{figure}

\newpage

\section{Discussion }

In the process of derivation for the formula, there are two
approximations to be used. One is the tight-bind approximation where
the pair-distribution function $g(\vec{r},\vec{r'})$ only involves
diagonal elements $\rho_{ii}$ of charge density. Other is that the
electron-electron interaction $v(\vec{r}-\vec{r'})$ is drawn out of
the integrand of the integral expression for $E_c$ (See
(\ref{eq:Ec-rr-2})), which was proved to be available in calculating
the excitation of the conjugated polymer oligomer$^{33}$. In this
way, the correlation energy expression obtained is directly related
to the diagonal site charge density
$\rho_{ii}^{\sigma}=<a^{\dag}_{i\sigma}a_{i\sigma}>$ and square of
the bond charge density
$\rho^{\sigma}_{ik}=<a^{\dag}_{i\sigma}a_{j\sigma}>$ ($i\neq k$)
that can be calculated directly from the elliptic integrals
(\ref{eq:first-elliptic}) and (\ref{eq:second-elliptic}). This is
different from the variational methods$^{22,23,24}$, and also from
the the density matrix renormalization group (DMRG)$^{34}$.

If the overlap effect of the $\pi$ electrons at adjacent sites is
included, the tight-bind approximation is invalid and the
pair-distribution function $g(\vec{r},\vec{r'})$ will have both the
diagonal terms $\rho_{ii}$ and the off-diagonal terms $\rho_{ij}$ of
the charge density. In this case, the electron charge density
$\rho_{ii}$ on the sites will be less than half and the bond charge
density $\rho_{ij}$ ($i\neq j$) will be enhanced a little. It may be
predictable that when the overlap effects of the $\pi$ electron
waves are considered, the correlation energies will become smaller.
Moreover, when the overlap integrals are considered, we may discuss
the contributions of the correlation energy from the off-diagonal
electron-electron interactions$^{35}$.

It is seen from Table 2 that for an infinite polymer PE, the
correlation energy was obtained and was $\epsilon_c(U+V)=-0.75$
eV$^{22}$. According to Ref.22, $77\% $ the correlation energy comes
from the contribution of the operator
$n_{i\uparrow}n_{i\downarrow}$. Thus the remaining 23$\%$ of the
correlation energy comes from the contribution of the long-range
Coulomb interaction $V_{ij}$ ($i\neq j$) or equivalently from the
contribution of the operator $n_in_j$ ($i\neq j$). From this, we may
have $\epsilon_c(V)=0.23\times (-0.75)=-0.1725$ eV. The present
correlation energy ($\epsilon_c(V)$) comes from the contribution of
the nearest-neighbor Coulomb interaction ($v=2.4$ eV) between two
$\pi$ electrons and is $-0.1696$ eV. In the Ref.23, the long-range
interaction $V$ (here $V=V_{i,i+1}=v$) was included within an
"effective $U$" after using some approximation, and the correlation
energy $\epsilon_c(U+V)$ both including the on-site Hubbard
interaction $U$ and the long-range Coulomb interaction $V$ and
$\epsilon_c(U)$ were obtained for PA, then the correlation energy
$\epsilon_c(V)$ from the contribution of the long-range Coulomb
interaction $V$ may be obtained from the difference between
$\epsilon_c(U+V)$ and $\epsilon_c(U)$ and it yielded $-0.1567$ eV
with $t_0=2.9$ eV. This value is smaller than the result of the
present study. The Ref.24 used the local approach$^{36}$ to
calculate the correlation energies. However, the difference between
$\epsilon_c(U+V)$ and $\epsilon_c(U)$ was very small and
$\epsilon_c(V)=-0.0100$ eV with $t_0=2.5$ eV in Ref.24, and this
result seems so small.

It is also seen from Table 2 that the present result of the
correlation energy $\epsilon_c(V)$ for PA is little smaller than
that for PE. Both PE and PA includes $sp^3$ hybridization. In the
calculation of the correlation energy for PE $^{22}$, except $\pi$
electrons between two nearest-neighbor carbon atoms , factors from
the different $\pi$ bonds were also taken in numerical computation,
so the more correlation effects were included in $\epsilon_c(U+V)$
for PE. In addition, the present result $\epsilon_c(V)=-0.1696$ eV
is little larger compared with $\epsilon_c(V)=-0.1567$ eV from
Ref.23. This may be caused by the approximation where the overlap
effect between two adjacent $\pi$ electronic wave functions is
omitted. When the overlap effect is considered in calculation, the
bond charge density will be larger and the site charge density will
be less than half, then the result will become smaller.

It is seen from (\ref{eq:Ec-9}) that the correlation energy is an
even function of dimerization. The trend of the curve in Fig.1 is
kind of quadratic but not complete because there are also the
dimerization parameter $z$ in the denominators in the formula. This
point can be seen by the following way. Because $z\ll 1$, if we
replace $z$ in the denominator in the formula by $z_0(=\delta
t_0/t_0)$, then the correlation energy $\epsilon_{c}$ is
approximately proportional to $-0.125
v+\frac{Av}{2\pi^2}+\frac{Bv}{2\pi^2}z^2$, where $A$ and $B$ are the
integral constants. At present, there is no similar curve to
compare. We may compare the curve in Fig.1 with those in Refs.24 and
27. Because the correlation energy (\ref{eq:Ec-9}) do not contain
the contribution form the on-site Hubbard interaction $U$, the curve
in Fig.1 is not completely like those in Fig.2 in Ref.24 and those
in Fig.1 in Ref.27 where both $U$ and the long-range interaction
were included. Nevertheless, it could be found here that the curve
in Fig.1 still have some similar trend with them when the
dimerization value is bigger. Though the curve in Fig.2 in Ref.27
was about the ground state energy vs the dimerization, the curve
also reflects some information about the correlation energy vs the
dimerization because the ground state energy contains the
correlation energy contribution in Ref.27.

It is all known that when screening is weak or normal in the $\pi$
electronic conjugated polymers, the electron-electron interaction
increases the dimerization and band energy gap$^{35}$. That is, $v$
increases $\delta t_0$ and $E_g$. In equilibrium state and rigid
background, the band energy gap $E_g$ of PA is given by $4\delta
t=4(\delta t_0+v\delta\rho_0)$. Under the electron-electron
interaction $v$, the average bond charge density $\bar\rho_0$
decreases slightly with increasing $\delta t_0$ and the fluctuation
of the bond charge density $\delta\rho_0$ increases with increasing
$\delta t_0$. The decrease of $\bar\rho_0$ causes the bandwidth (see
$t=t_0+v\bar\rho_0$) to diminish, and the increase of $\delta\rho_0$
make $E_g$ increase and at the same time cause $\epsilon_c$ (see
(\ref{eq:Ec-8})) to decrease. Then it can be seen from these that
$v$ and $\delta\rho_0$ are two opposite factors to the correlation
energy: $v$ is in favor of the correlation energy but $\delta\rho_0$
is a disadvantage to the correlation energy in this study. That is
to say, on the one hand, $v$ makes $\delta t_0$ and accordingly
$\delta\rho_0$ increase and then the band energy gap $E_g$ increase,
on the other hand, the fluctuation $\delta\rho_0$ will cause the
correlation energy $\epsilon_c$ to decrease. As a result, these two
opposite influence makes $E_g$ decrease from $E_g=4\delta t$ to
$E'_g=E_g+\epsilon_c$. When $\delta t_0=0.269$ eV and the
corresponding bond charge density $\delta\rho_0=0.0903$, we have
$E_g=1.9429$ eV and $\epsilon_c=-0.1696$ eV. Therefore the band
energy gap $E'_g$ containing the nearest-neighbor Coulomb
interaction correlation effect becomes $1.7832$ eV. This value of
the band energy gap is close to $E''_g$ ($1.8$ eV) obtained by $Ab$
$initio$ computation by author in Ref.27 where the screened
interaction was used. These are in qualitative agreement with
experiment. In addition, $E'_g$ increase with increasing
dimerization because the correlation energy $\epsilon_c$ decreases
with dimerization, which is consistent with relation of the band gap
and the dimerization.

To my knowledge, there is no similar expressions for the correlation
energy only from the nearest-neighbor Coulomb interaction that
exists in a simple form at present. In the next study, the more
electron-electron interaction terms in the long-range Coulomb
interaction will be considered. In addition, although the DMRG is a
strong tool to deal with the correlation problems in a many-particle
electron system, it is basically a complicated numerical calculation
method but not is an analytical expression.

$In$ $summary$, under the approximation (\ref{eq:Ec-rr-2}) and the
tight-bind approximation, a formula (\ref{eq:Ec-6}) of the
correlation energy for the long-range (nearest-neighbor) Coulomb
interaction $v$ for the conjugated polymers is obtained with the
rigid backbone background. Although it is simple, it is direct and
effective and easily operational in comparison with other highly
involved numerical computation methods including DMRG. The
computational result for the correlation energy for PA is available
compared to those for PA and PE in different methods$^{22,23}$. The
band energy gap $E'_g$ containing the correlation effect is close to
that by $ab$ $initio$ method containing the screening
interaction$^{27}$. Because there are no constraints to the systems
in the process of the deduction, this relation may be applied to the
various $\pi$ electron systems such as C$_{60}$, benzene rings, and
carbon nanotubes, etc.

\begin{center}
\large {\bf ACKNOWLEDGMENT}
\end{center}
Special Research Fund (2005) of ChongQing University, ChongQing,
P.R. China is acknowledged.

\end{document}